# Fully achromatic nulling interferometer (FANI) for high SNR exoplanet characterization

François Hénault
Institut de Planétologie et d'Astrophysique de Grenoble
Université Joseph Fourier, Centre National de la Recherche Scientifique
B.P. 53, 38041 Grenoble – France

## ABSTRACT

Space-borne nulling interferometers have long been considered as the best option for searching and characterizing extra-solar planets located in the habitable zone of their parent stars. Solutions for achieving deep starlight extinction are now numerous and well demonstrated. However they essentially aim at realizing an achromatic central null in order to extinguish the star. In this communication is described a major improvement of the technique, where the achromatization process is extended to the entire fringe pattern. Therefore higher Signal-to-noise ratios (SNR) and appreciable simplification of the detection system should result. The basic principle of this Fully achromatic nulling interferometer (FANI) consists in inserting dispersive elements along the arms of the interferometer. Herein this principle is explained and illustrated by a preliminary optical system design. The typical achievable performance and limitations are discussed and some initial tolerance requirements are also provided.

## 1 INTRODUCTION

Nulling interferometry is probably the most promising technique for searching and characterizing habitable extra-solar planets in the next decades. Starting from the original concept proposed by Bracewell [1] and pioneering works of Léger *et al* [2] or Angel and Woolf [3], it led to detailed industrial studies for the realization of two futuristic space interferometers, namely Darwin for the European Space Agency [4] and the TPF-I for US National Aeronautics and Space Administration [5]. During those years the major technical challenges were overcome, including manufacturing of mid-infrared waveguides for wavefront error filtering [6], validations of different types of Achromatic phase shifters (APS) generating the dark central fringe [7-8], and laboratory demonstrations of the required extinction performance over a wide spectral range [9-10]. Today the sole obstacle that remains ahead nulling interferometry (and more generally interferometric imaging in space) seems to be the accurate control of free-flying space vessels.

Surprisingly, and while realizing deep achromatic central nulls is now well-mastered, it seems that little attention was paid to achromatizing the entire fringe pattern, which should result in higher Signal-to-noise ratios (SNR) and appreciable simplification of the detection system. In this communication are presented the general principle of a Fully achromatic nulling interferometer (FANI) based on the insertion of dispersive elements along the interferometer arms. Herein this principle and its governing mathematical relationships are described in section 2. A tentative optical design and its expected performance are presented in section 3, including a methodology for dispersive optics dimensioning (§ 3.1), numerical simulations of the generated fringe patterns for different interferometric configurations (§ 3.2), and a preliminary ray-tracing optical design and tolerance analysis (§ 3.3). Finally, a few words about specific advantages (high SNR planet detection, potential use as a broadband imaging interferometer) and a short conclusion are provided in sections 4 and 5 respectively.

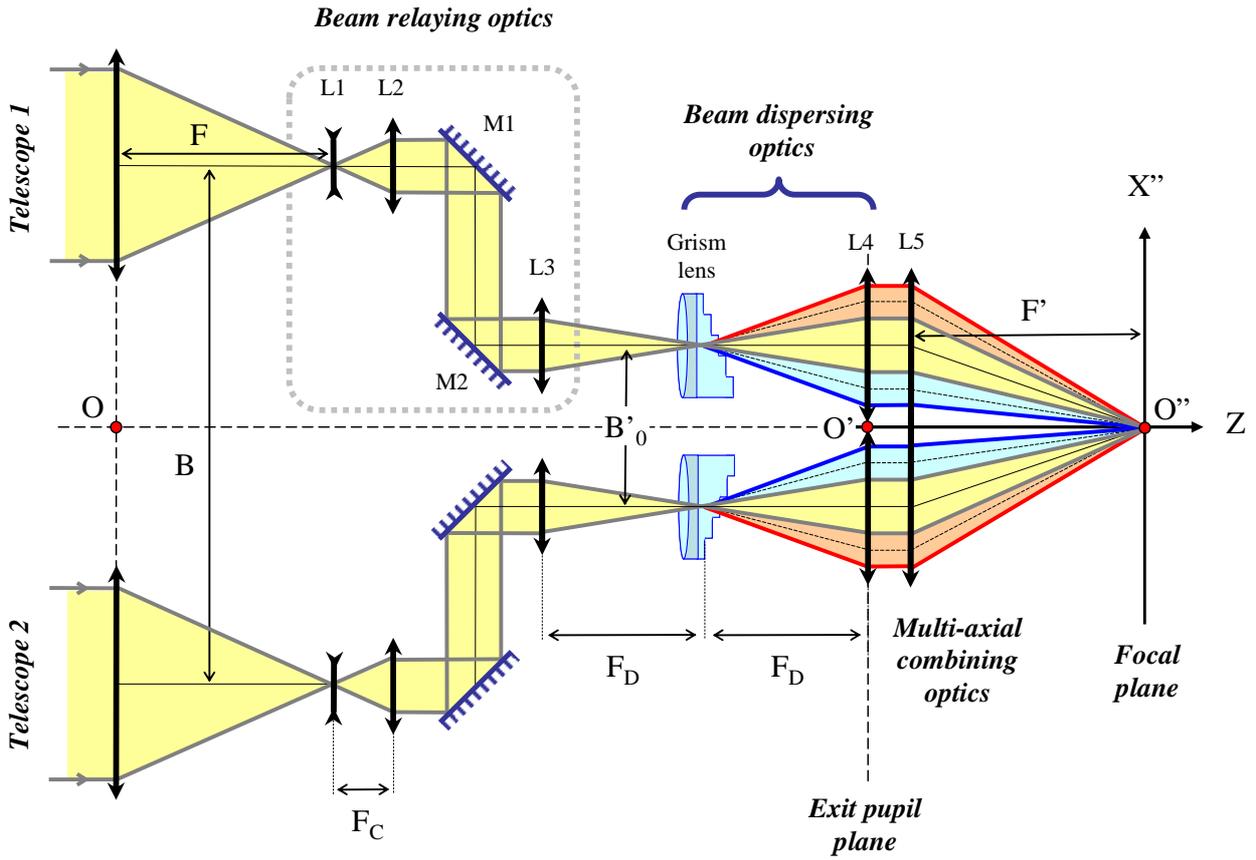

Figure 1: .General optical scheme of the achromatic nulling interferometer.

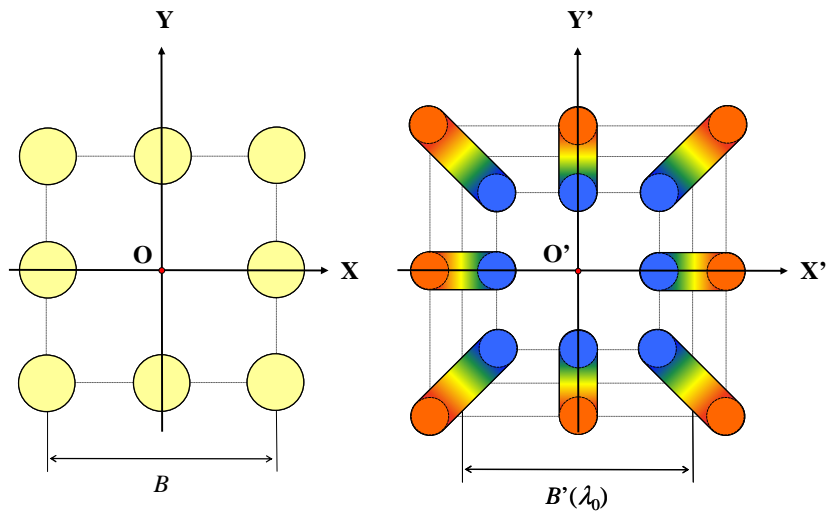

Figure 2: Sketch of the input and output sub-apertures of an eight-telescope interferometer, respectively on left and right panels. The main entrance baseline is noted B and the exit baseline B' is a function of the wavelength λ. Blue and red-colored exit pupils respectively correspond to the shortest and longest wavelengths noted $\lambda_B$ and $\lambda_R$. $\lambda_0$ designates an average reference wavelength.

## 2  GENERAL PRINCIPLE

This section includes a general description of the achromatic nulling interferometer (§ 2.1) and simplified theoretical expressions of the produced fringe patterns (§ 2.2).

### 2.1  General description

The basic principle of the Fully achromatic nulling interferometer (FANI) is depicted in Figure 1 and Figure 2, referring to coordinate systems and scientific notations defined into the Appendix. We should firstly note that it is only applicable to interferometers equipped with multi-axial combining optics, that is fortunately the case of most existing facilities. Provided that this necessary condition is fulfilled, the FANI is composed of the main following optical components or sub-systems:

- The collecting telescopes whose primary mirrors are assumed to be the effective entrance pupils of the interferometer arms. Their diameter and focal length are denoted D and F respectively.

- The Beam relaying optics (BRO), fulfilling the double function of conveying the images from the telescope to the detector plane, and of optically conjugating the telescope entrance pupil with the exit pupil of the interferometer that is located on the multi-axial combiner (see below). It includes various optical components indicated as thin lenses in Figure 1, such as collimating optics L2, refocusing optic L3 and pupil imaging optic L1 (note that its representation as a diverging element in the figure is only indicative). Here the distance between mirrors M1 and M2 is variable and adjusted to fit the entrance baseline B of the interferometer. Together with F, the entrance focal length $F_C$ of the BRO determines the optical compression factor of the interferometer, equal to $m = F_C/F$.

- The Beam dispersing optics (BDO). These are the core components of FANI. They are essentially composed of a dispersing element located at the focus of L3 (named grism lens in Figure 1), and of a collimating lens L4 of focal length $F_D$. The main function of the grism element is to deviate the incoming beam under a given deviation law $\beta(\lambda)$ as function of the incident wavelength, therefore turning the output baseline B' of the interferometer into a varying function B'($\lambda$). It is then assumed that:
   1- The central wavelength $\lambda_0$ of the useful spectral band is not deviated by the grism, thus defining a reference baseline $B'_0$ in the exit pupil plane of the interferometer
   2- The grism is dimensioned so as to generate an output baseline function $B'(\lambda) = B'_0 \lambda/\lambda_0$.

Therefore the natural $1/\lambda$-chromatism induced by diffraction should be compensated for, and a fully achromatized fringe pattern will result in the focal plane of the interferometer. This spectral pupil remapping technique is schematically illustrated in Figure 2 and Figure 3. Obviously both previous conditions can only be met with careful matching of the $\lambda_0$, $B'_0$ and $F_D$ parameters, as will be explained in section 3.1.

- The Multi-axial combining optics (MCO) located just after the BDO, that collect its output collimated beams and focuses them at the image plane of the interferometer, where the achromatic fringe pattern can be observed on a detector array.

- Finally, an Achromatic phase shifter (APS, not shown in the Figures) must be incorporated into the system in order to transform the naturally bright central fringe of the interferometer into a dark one, suitable to star extinction and exoplanet characterization. Among the current variety of existing APS types, we should select here the dispersive plate concept [8-9] because dispersive materials are already employed into the BDO and potentially usable for global optimization.

Single-mode waveguides are also necessary for filtering the Wavefront errors (WFE) of the whole optical train and achieving deep extinctions rates. They generally are located downstream the focal plane and are not discussed in this communication.

### 2.2  Theoretical relationships

Before dimensioning and analyzing a real optical system in section 3, it is worth giving here simplified theoretical expressions of the fringe patterns generated by FANI. The following expressions are derived from a mathematical formalism already described in [11] and some of the cited references. It is summarized into the Appendix, providing a short but complete description of useful coordinate frames and mathematical notations. From Eq. A4 in the Appendix and setting the angular coordinates $u_0 = v_0 = 0$, we find that the monochromatic fringe pattern, or transmission map $T_\lambda(u,v)$ at the centre of the interferometer Field of view (FoV) writes as:

$$T_\lambda(u,v) = \left|\frac{2J_1(\pi Dr/\lambda)}{\pi Dr/\lambda}\right|^2 \left|\sum_{n=1}^{N} a_n \exp[i\varphi_n] \exp\left[-i\frac{2\pi}{\lambda}\left(\frac{ux'_n + vy'_n}{m}\right)\right]\right|^2, \quad (1)$$

where $r = \sqrt{u^2 + v^2}$ and the meaning of all other mathematical symbols is explicated in the Appendix. Here it is more convenient to reduce this very general relationship to the simpler case of a N = 2-telescope interferometer of entrance baseline B, therefore becoming:

$$T_\lambda(u,v) = \left|\frac{2J_1(\pi Dr/\lambda)}{\pi Dr/\lambda}\right|^2 \sin^2\left(\frac{\pi}{\lambda}\frac{B'(\lambda)u}{m}\right), \quad (2)$$

with $B'_0$ the exit baseline of the interferometer at the reference wavelength $\lambda_0$; as introduced is the previous sub-section. Assuming now that dispersive optics obey to the law of diffraction gratings expanded at the first-order (i.e. diffracted angles $\beta$ being proportional to the wavelength $\lambda$), and that they were dimensioned to have $B'(\lambda)$ equal to $B'_0 \lambda/\lambda_0$, Eq. 2 subsequently leads to:

$$T_\lambda(u,v) = \left|\frac{2J_1(\pi Dr/\lambda)}{\pi Dr/\lambda}\right|^2 \sin^2\left(\frac{\pi}{\lambda_0}\frac{B'_0 u}{m}\right). \quad (3)$$

It can be seen that the spectral dependence in $\lambda$ has been removed from the fringe pattern term (the rightmost on the right-hand side of Eq. 3), and is only remaining in the fringe envelope term $E(u,v) = \left|2J_1(\pi Dr/\lambda)/\pi Dr/\lambda\right|^2$ (the leftmost term). The resulting polychromatic, wideband interferogram can now be computed by numerically integrating Eq. 3 over the useful spectral range of the interferometer, that may be set by detector bandwidth, dichroïc coatings, spectral filters or any other similar item. Denoting $\lambda_B$ as the shortest or "blue" wavelength, $\lambda_R$ the longest or "red" wavelength, and assuming a uniform spectral distribution between $\lambda_B$ and $\lambda_R$, this leads to:

$$T(u,v) = \int_{\lambda_B}^{\lambda_R} T_\lambda(u,v) \, d\lambda = \sin^2\left(\frac{\pi}{\lambda_0}\frac{B'_0 u}{m}\right) \int_{\lambda_B}^{\lambda_R} \left|\frac{2J_1(\pi Dr/\lambda)}{\pi Dr/\lambda}\right|^2 d\lambda. \quad (4)$$

The result is equal to the product of the monochromatic fringe pattern at the reference wavelength $\lambda_0$ with a spectrally averaged envelope function of approximate angular radius $\lambda_0/D$, and where the external diffraction rings in function $E(u,v)$ should be washed out by the integrating operation.

From the previous theoretical analysis, it follows that if equipped with the FANI combining optics, any type of interferometer (i.e. whatever is the geometrical arrangement of its entrance telescopes) can be used to generate an achromatic fringe pattern at its FoV centre. It must be noted however that there will remain a chromatic dependence in the interferometer FoV due to the $2\pi(u_0 x_n + v_0 y_n)/\lambda$ terms in Eq. A4. This dependence could also be removed by applying a similar, but reversed achromatizing process to the input optics of the interferometer. However this would probably require much more complicated optical designs.

## 3    DESIGN AND PERFORMANCE

Following the previous theoretical analysis and sub-systems definition, here are presented the methodology for dimensioning the dispersive optics (§ 3.1), numerical simulations of the created fringe patterns for different sub-aperture arrangements (§ 3.2), and a preliminary ray-tracing optical design and tolerance analysis of the system (§ 3.3).

## 3.1 Dispersive optics and grism definition

As explained in § 2.1, the basic principle of FANI consists in setting a dispersive element at an intermediate image plane of the system, an idea that was originally proposed for designing integral field spectrographs [12] or focal plane wavefront sensors [13], but not for an entire interferometer. For this last application the chosen dispersing element will be a grism associated with a collimating lens L4, together forming the BDO sub-system. The general approach for dimensioning the BDO is illustrated in Figure 3 and explained below.

We start from the classical grism equation that writes:
$$k\,\lambda = a\,[n(\lambda)\sin\alpha - \sin(\alpha + \beta(\lambda))], \tag{5}$$

where  $k$  is the diffraction order,
$a$  is the spatial period of the diffraction grating (see Figure 3),
$n(\lambda)$  is an analytic function defining the refractive index of the prism material as function of wavelength,
$\alpha$  is the prism angle,
$\beta(\lambda)$  is the angle between the diffracted ray and the optical axis, also depending on wavelength.

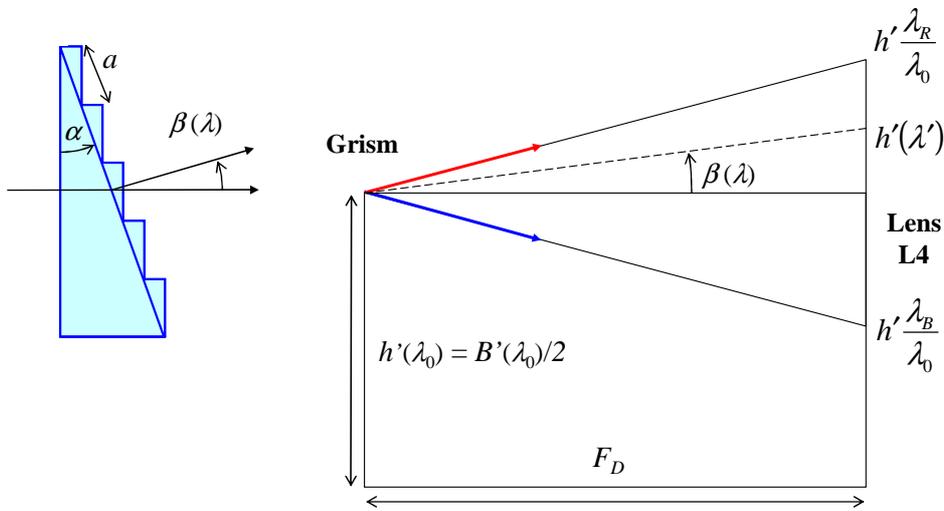

Figure 3: Illustrating the main dispersive optics and grism parameters.

Selecting the grating order $k = 1$ and assuming that its peak wavelength coincides with $\lambda_0$ at $\beta = 0$, Eq. 5 leads to:
$$\lambda_0 = a\,[n(\lambda_0) - 1]\sin\alpha\ . \tag{6}$$

Secondly, the rays diffracted by the grism at an height $h' = B'_0/2$ shall impact the L4 plane at the height $h'\lambda/\lambda_0$, a condition that can be written as:
$$h' + F_D \tan[\beta(\lambda)] = h'\frac{\lambda}{\lambda_0} + dh'(\lambda); \tag{7}$$

where $dh'(\lambda)$ is an error term quantifying the distortion with respect to the ideal linear deflection law. In the first-order Gaussian approximation domain ($\beta(\lambda) \ll 1$) and using the sign conventions in Figure 3, Eqs. 5 and 7 can respectively be expressed as follows.

Grism equation:
$$\beta(\lambda) \approx -\frac{\tan\alpha}{\lambda_0}\{\lambda_0\,n(\lambda) - \lambda\,n(\lambda_0) + \lambda - \lambda_0\}, \tag{8}$$

Ray impact on L4: $$\beta(\lambda) \approx \frac{h'}{F_D}\frac{\lambda-\lambda_0}{\lambda_0} + \frac{dh'(\lambda)}{F_D}. \qquad (9)$$

It is also assumed that the refractive index function $n(\lambda)$ can be developed as:

$$n(\lambda) = n_0 + \nu_0(\lambda-\lambda_0) + dn(\lambda), \qquad (10)$$

with $n_0 = n(\lambda_0)$, $dn(\lambda_0) = 0$. Here $\nu_0$ stands for the spectral slope of the refractive material at the first-order, and $dn(\lambda)$ represents higher-order terms. Both quantities can be computed numerically from the Sellmeier law of the employed material averaged over $[\lambda_B, \lambda_R]$, the useful spectral range of the interferometer. Equating now relations 8 and 9 term by term finally gives suitable expressions for estimating the grism angle $\alpha$ and spectral distortion relative errors $dh'(\lambda)/h'$:

$$\tan\alpha = \frac{h'}{F_D(n_0-1-\nu_0\lambda_0)}, \quad \text{and:} \qquad (11)$$

$$\frac{dh'(\lambda)}{h'} = \frac{dn(\lambda)}{n_0-1-\nu_0\lambda_0}. \qquad (12)$$

It can be noticed that the grism angle $\alpha$ only depends on the ratio $h'/F_D$, and that the spatial period $a$ of the diffraction grating is easily obtained by inverting the grism equation 6, thus giving $a = \lambda_0/[n(\lambda_0)-1]\sin\alpha$. For a $h'/F_D$ ratio equal to 1/5 as in the following sub-section, the grism parameters and achievable performance in terms of distortion are then estimated for various typical infrared materials listed in Table 1.

Table 1: Grism parameters and performance for various typical infrared materials.

| Material | Refractive index at $\lambda_0$ | Spectral slope $\nu_0$ (µm$^{-1}$) | Grism angle (°) | Groove period (µm) | RMS distortion (%) |
|---|---|---|---|---|---|
| CdTe | 2.700 | -2.7E-03 | 6.710 | 53.77 | 2.1E-02 |
| CsI | 1.747 | -8.2E-04 | 14.982 | 54.98 | 1.7E-02 |
| KBr | 1.548 | -2.3E-03 | 20.018 | 58.40 | 7.8E-02 |
| KCl | 1.496 | -4.0E-03 | 21.934 | 61.85 | 1.6E-01 |
| KRS5 | 2.392 | -2.2E-03 | 8.174 | 53.92 | 2.2E-02 |
| NaCl | 1.526 | -1.2E-05 | 20.802 | 56.17 | 1.3E-03 |
| ZnS | 2.336 | -1.4E-02 | 8.488 | 59.69 | 2.6E-01 |
| ZnSe | 2.470 | -6.5E-03 | 7.741 | 55.55 | 9.2E-02 |

## 3.2 Simulated fringe patterns

From the theoretical relationships presented in sub-sections 2.2 and 3.1, it becomes possible to perform more realistic numerical simulations of the fringe patterns generated by FANI, i.e. taking into account the effective indices of the employed materials. The choice of this material shall obviously be based on the data of Table 1, but also on practical considerations about manufacturing, environment constraints, and space qualification of the candidate materials. Here we finally selected ZnSe, assumed to be a good compromise between performance and realization difficulties.

The main physical and geometrical parameters of the interferometer are summarized In Table 2. It must be noted that the studied configurations correspond to an ideal Fizeau interferometer at the reference wavelength $\lambda_0$ (i.e. the entrance and exit pupils are perfectly homothetic at $\lambda_0$, see the Appendix). However this condition is not restrictive since the FANI combining optics arrangement is independent of the entrance pupil configuration as mentioned in § 2.2. For a N-telescope interferometer and using the same mathematical formalism, a polychromatic interferogram generated by FANI can now be computed numerically from the integral:

$$T(u,v) = \int_{\lambda_B}^{\lambda_R} T_\lambda(u,v)\, d\lambda = \int_{\lambda_B}^{\lambda_R} \left| \frac{2J_1(\pi D r/\lambda)}{\pi D r/\lambda} \right|^2 \left| \sum_{n=1}^{N} a_n \exp[i\varphi_n] \exp\left[ -i\frac{2\pi}{\lambda}\left( \frac{u x'_n(\lambda) + v y'_n(\lambda)}{m} \right) \right] \right|^2 d\lambda, \quad (13)$$

where functions $x'_n(\lambda)$ and $y'_n(\lambda)$ are the spectral coordinates of the $n^{th}$ exit sub-pupil, schematically represented in Figure 2 and having the following expressions:

$$x'_n(\lambda) = x'_n(\lambda_0)\left[1 + \frac{F_D \tan[\beta(\lambda)]}{\sqrt{x'_n(\lambda_0)^2 + y'_n(\lambda_0)^2}}\right] \quad \text{and:} \quad y'_n(\lambda) = y'_n(\lambda_0)\left[1 + \frac{F_D \tan[\beta(\lambda)]}{\sqrt{x'_n(\lambda_0)^2 + y'_n(\lambda_0)^2}}\right]. \quad (14)$$

The angle $\beta(\lambda)$ in Eqs. 14 is computed from inverting the grating equation 5, using the real refractive index function $n(\lambda)$ of ZnSe material and the $a$ and $\alpha$ parameters indicated in Table 1. In that case the theoretical fringe patterns formed by a N=2, 4 and 8-telescope nulling interferometer are reproduced in Figure 4, showing the monochromatic interferograms at $\lambda_0 = 10.5$ µm, the uncompensated broadband interferograms between $\lambda_B = 7$ and $\lambda_R = 14$ µm, and those generated by FANI on the same spectral width. The comparison reveals no perceptible difference between the purely monochromatic and achromatized fringe patterns. For the two-telescope case, this is confirmed by their cross-sections along the X'-axis depicted in Figure 5, where only small differences never exceeding 3 % can be observed near the fringes maxima. One can finally remark that FANI is producing slightly "colored' fringe images in the sense that only the longest (or "reddest') wavelengths can be observed at the edges of the integrated envelope function E(u,v) in Eqs. 3 and 4, therefore rendering the outermost diffraction lobes in red.

Table 2: Main parameters employed for numerical simulations.

| PARAMETERS | NUMERICAL VALUES |
|---|---|
| Number of entrance and exit pupils | N = 2, 4 and 8 |
| Main reference wavelength | $\lambda_0$ = 10.5 µm |
| Spectral range of the interferometer | $[\lambda_B, \lambda_R]$ = [7,14] µm |
| Main entrance baseline along X and Y axes | B = 20 m |
| Diameter of individual telescopes (entrance sub-pupil) | D = 5 m |
| Telescope focal length | F = 50 m |
| BRO focal length | $F_C$ = 10 mm |
| BRO compression factor | m = 1/500 |
| Dispersive optics focal length | $F_D$ = 100 mm |
| Main exit baseline along X' and Y' axes at the reference wavelength $\lambda_0$ | B'$_0$ = 40 mm |
| Diameter of exit sub-pupils | D' = 10 mm |

### 3.3 Optical design and tolerances

After first-order dimensioning of the dispersive optics sub-system in the previous sub-sections, it is now feasible to implement a preliminary optical design and to optimize it with the help of a ray-tracing software (herein Zemax). For now our main purpose is not to define a complete optical description of the optical layout, but rather to identify the main technical difficulties and giving hints to solve them.

*BDO optical design*

A preliminary optical design of the FANI dispersive optics is schematically illustrated in Figure 6 and Figure 7 (only one interferometer arm is shown, the others – apart from the APS – being symmetrical with respect to the X'Z or Y'Z planes). Excepting the grism, it is based as much as possible on reflective optics and comprises the following components:

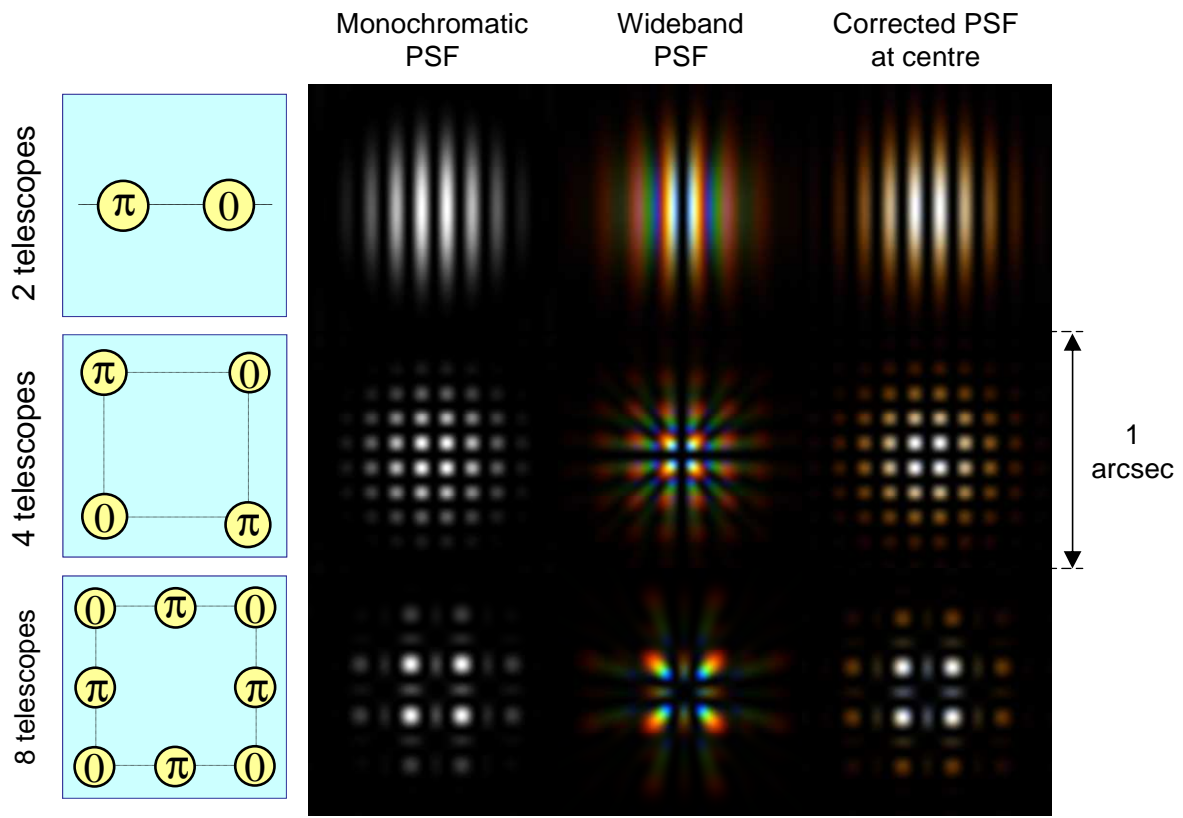

Figure 4: Theoretical fringe patterns generated an eight, four and two-telescope nulling interferometer. From left to right are shown the entrance pupils maps and their achromatic phase-shifts $\varphi_n$, the monochromatic interferograms at $\lambda_0 = 10.5$ µm, the uncompensated wideband interferograms between 7 and 14 µm, and those generated by FANI on the same spectral band.

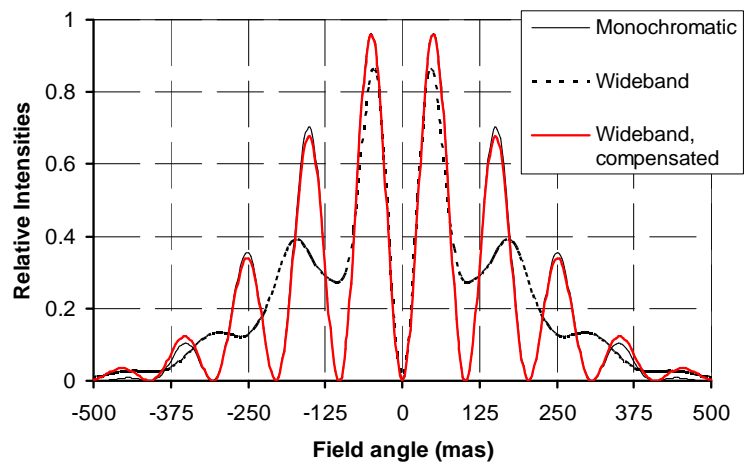

Figure 5: .Fringe patterns generated by a two-telescope interferometer. Thin black line: monochromatic interferogram at $\lambda_0 = 10.5$ µm. Dashed line: broadband interferogram between 7 and 14 µm. Red solid line: achromatic interferogram generated by FANI on the same wave band. The latter can hardly be distinguished from the monochromatic case.

- Following the path of incoming photons, the first encountered optical element is the dioptric APS whose detailed characteristics are given in the next paragraph.

- The focusing optics L3 and re-collimating optics L4 are in first approximation the off-axis sections of a common parabolic mirror of focal length $F_D$ = 100 mm. Their shapes will be slightly modified during the optimization process, as explained later.

- A reflective ZnSe grism is located near the common focus of the parabolic mirror sections. The thickness at centre of the grism is 2 mm and its other parameters are those defined on the last row of Table 1. Though not critical because the interferometer FoV is very small as mentioned in § 2.2, the curvature of the back reflecting face has been optimized so as to conjugate an entrance pupil located at -150 mm from the focusing mirror onto the combining mirror of the interferometer. The resulting curvature radius was found around -340 mm.

- Also shown in the figures but not part of dispersive optics sub-system, all interferometer beams are finally combined multi-axially by a parabolic mirror of focal length F' = 1 m located at 200 mm from the BDO collimating mirror.

The next step of the design process consists in optimizing the shape of the off-axis parabolic mirrors L3 and L4 in order to achieve a diffraction-limited image quality on the complete spectral band, while maintaining a negligible lateral chromatism at the interferometer focus. For this two options can be envisaged:

1. Either using a deformable mirror (DM) at the location of L4 as sketched in Figure 6. This configuration is somewhat similar as that employed for the adaptive nuller experiment [14], with the difference that the locations of the DM and dispersive element have been exchanged. Here the DM is used for dynamically compensating chromatic off-axis aberrations generated by the grism mirror. However the technical realization of a cryogenic DM operating in the mid-infrared waveband may not be the simplest technical solution.

2. Either preferring a static collimating mirror for cost and development complexity reasons. In that case the aspheric coefficients of the focusing and collimating mirrors may be adjusted for OPD equalization and minimization of residual chromatic aberrations.

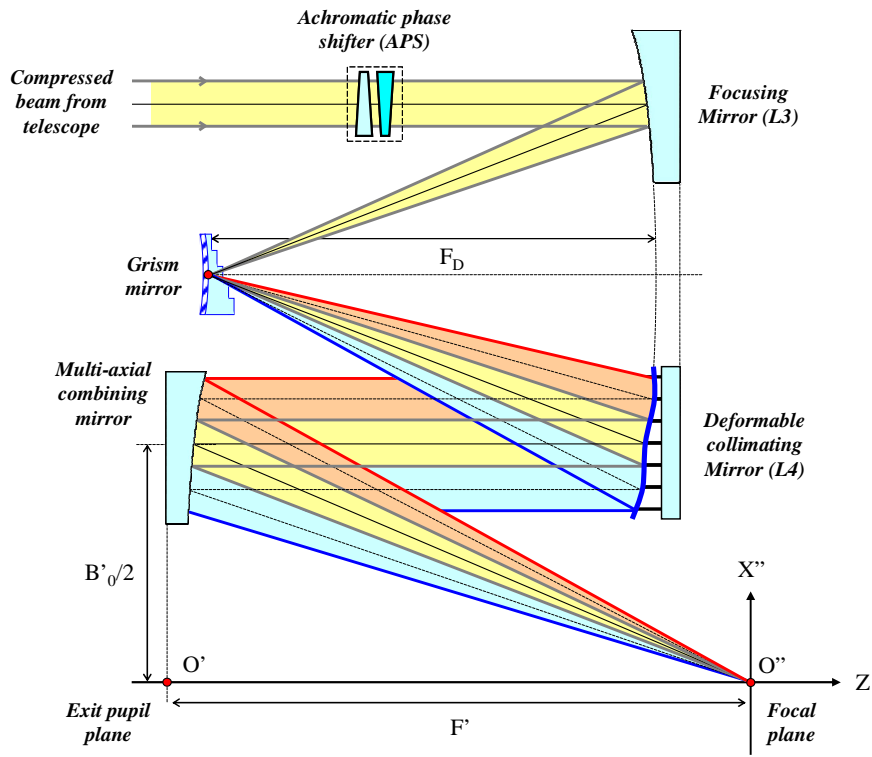

Figure 6: Possible optical design for dispersive optics subsystem, including the achromatic phase shifter, grism mirror and a collimating deformable mirror usable for OPD equalization. Only one interferometer arm is shown.

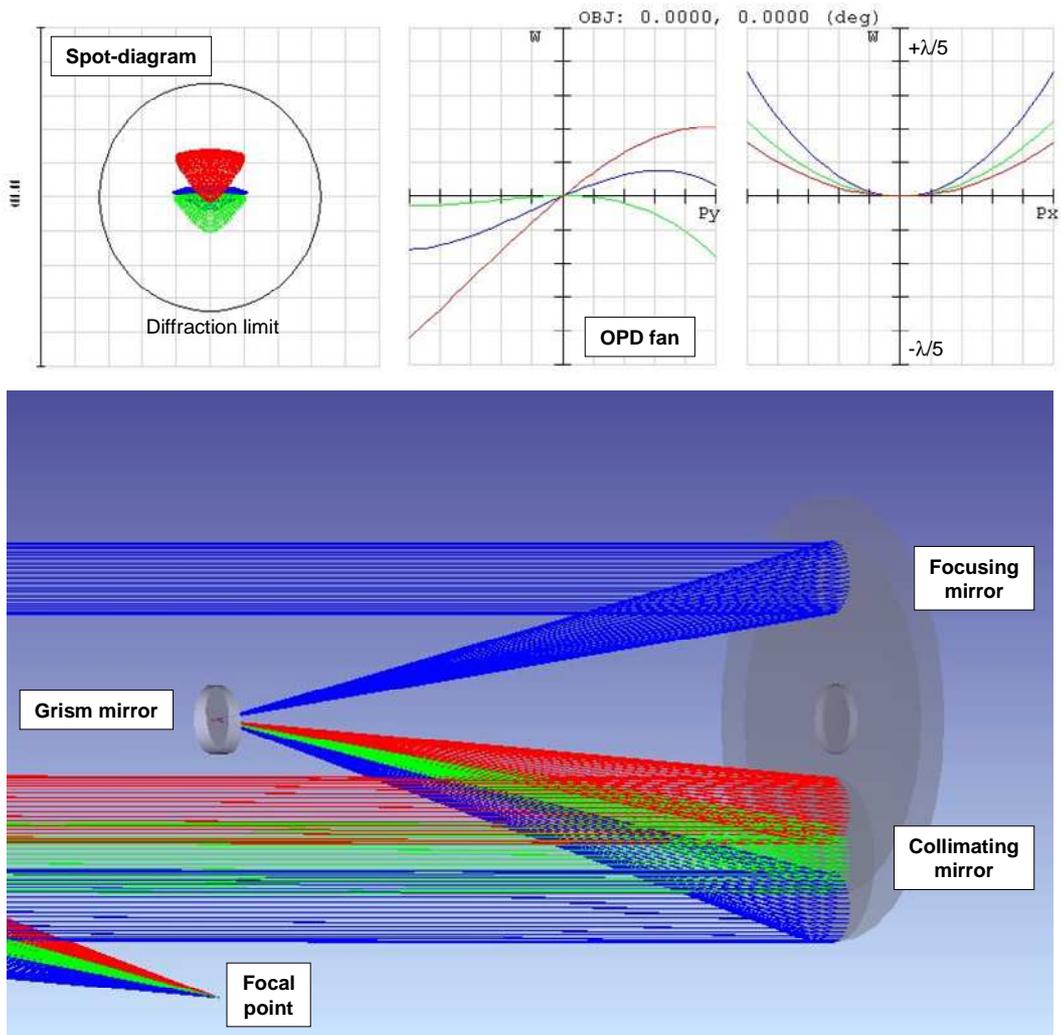

Figure 7: Zemax ray-tracing model of the dispersive optics (spot-diagram, OPD fans and 3D layout).

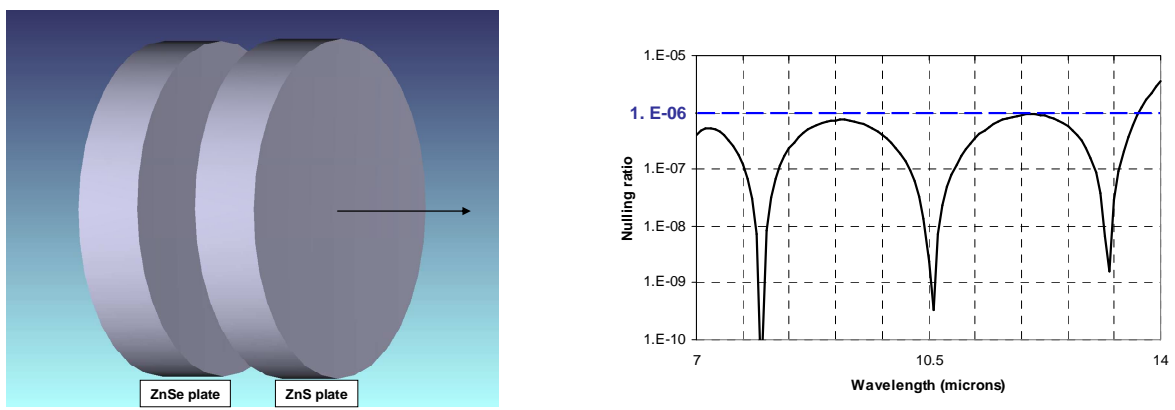

Figure 8: .View of the APS wedge plates and their performance in terms of nulling rate as function of wavelength (logarithmic scale).

In this study we opted for the second solution. Then the optimization process led to aspherization coefficients being equal to -2.225 (hyperbolic shape) and -0.775 (elliptic shape) respectively for the focusing and collimating mirrors. The attained performance is illustrated in Figure 7, showing spot-diagrams and OPD fans well below the diffraction limit. We also checked that the RMS wavefront error is always inferior to $\lambda/25$ waves on the full interferometer spectral range. It must be emphasized that the optimization was carried out in presence of the APS wedge plates, which are defined in the next paragraph.

*Achromatic phase shifter (APS)*

As already stated in § 2.1, the selected concept is the refractive plates or "dioptric" APS that was successfully validated on the MAII nulling test bench [8] and selected as one of the two phase-shifters implemented into the PERSEE experiment [9]. Its basic principle consists in installing a pair of wedge dispersive plates made of two different materials along each arm of the interferometer, and adjusting their individual thickness and wedge angle in order to:

- Ensure a constant, achromatic $\pi$ phase-shift between an interferometer arm and its symmetric counterpart,
- Maintain the parallelism of the exit beams before final combination.

Here ZnSe is an obvious choice for the first refractive material, since it is already present into the BDO grism mirror. We also select ZnS as the companion material and evaluated the four plate thicknesses in order to generate the achromatic $\pi$ phase-shift as explained in Ref. [9]. The APS and grism mirror were globally optimized together in order to cancel the lateral chromatism at the focus of the interferometer, or to minimize it below the diffraction limit (see the upper left panel of Figure 7). The derived optical characteristics of the APS are summarized in Table 3 and the attainable performance in terms of nulling rate vs. wavelength is illustrated in Figure 8. It can be seen that the null ratio remains lower than $10^{-6}$ over 95 % of the considered spectral band (a $10^{-6}$ nulling rate was the original specification for Darwin or TPF-I [4-5]). Finally, it should be noted that FANI could easily be turned into a classical stellar imaging interferometer generating a bright central fringe by simply removing the APS from the optical paths with the help of flipping or equivalent types of mechanisms.

Table 3: Optical characteristics of the achromatic phase-shifter. Plates 1' and 2' are the counterparts of plates 1 and 2 in the symmetric interferometer arm.

| Plate number | Material | Thickness (mm) | Wedge (degs.) |
|---|---|---|---|
| 1 | ZnS | 2 | 0.666 |
| 2 | ZnSe | 2 | 0.5 |
| 1' | ZnS | 2.202 | 0.666 |
| 2' | ZnSe | 1.482 | 0.5 |

Table 4: Preliminary BDO tolerance analysis.

| Geometrical parameter | Tolerance |
|---|---|
| Grism mirror translation along Z-axis | $\leq 0.1$ mm |
| Grism mirror decenter (along X' and Y' axes) | $\leq 1$ mm |
| Grism mirror tilt around X'-axis | $\leq 5$ degs. |
| Grism mirror tilt around Y'-axis | $\leq 1$ deg. |
| Grism mirror roll angle (around Z-axis) | $\leq 5$ degs. |
| Grism thickness at centre | $\leq 0.1$ mm |
| Grism angle $\alpha$ | $\leq 1$ deg. |

*Tolerance analysis*

Building a complete tolerance analysis of FANI including all BDO and APS components is obviously an important task that should be carried out in the future. It is however beyond the scope of this communication because it would necessitate developing more detailed models of the nulling ratio degradations on presence of optical defects. Moreover, previous nulling error budgets were already established on similar types of designs (see e.g. Ref. [15]). The present study is thus limited to the grism mirror, which is the core BDO component and should be manufactured and aligned within the diffraction limit of the collecting telescopes. Strictly complying with this criterion, the preliminary alignment and manufacturing tolerances for the BDO dispersive element are summarized in Table 4. Technically speaking, none of them can be considered as critical.

## 4 DISCUSSION

Because the past studies of Darwin and TPF-I were more focused at major critical issues such as wavefront filtering, or APS design [4-5], it seems that less attention was paid to the final detection stage, essentially composed of a low dispersion spectrograph fed by a Single-mode fiber (SMF), and of a detector array as illustrated on the left side of Figure 9. In this basic configuration, only one SMF can be placed at the FoV centre, that is the unique location where the planet signal (the blue curves in the Figure) emerges from the residuals of the extinguished star (red curves) due to chromatic dispersion of the interference patterns. Conversely, an interferogram generated by FANI offers a higher contrast at the FoV centre and several different locations where the SMF entrance tips can be positioned (right side of Figure 9). These fibers can then connected downstream to a long slit or multi-object spectrograph, enabling a significant SNR gain for extra solar planet detection and characterization. This potential advantage should however be demonstrated more rigorously in future work, for example using the same mathematical formalism as presented in Ref. [16] and applied to fibered nulling telescopes.

Though not illustrated here, FANI can also be operated as a classical stellar interferometer for broadband imaging purpose, where it offers a different advantage with respect to most other existing facilities. Actually these interferometers almost always incorporate dispersive elements (often diffraction gratings) for analyzing specific spectral lines mapped onto a bi-dimensional detector array. Broadband images are then reconstructed by adding together contributions from different spectral channels. For a $n$-channel spectrograph, readout and dark noises consequently add into a $\sqrt{n}$ proportion. Such noise accumulation effect should not occur on a FANI-like, wideband imaging interferometer where only a few detector rows are used, and a SNR gain of the order of $\sqrt{n}$ is to be expected with respect to usual interferometers.

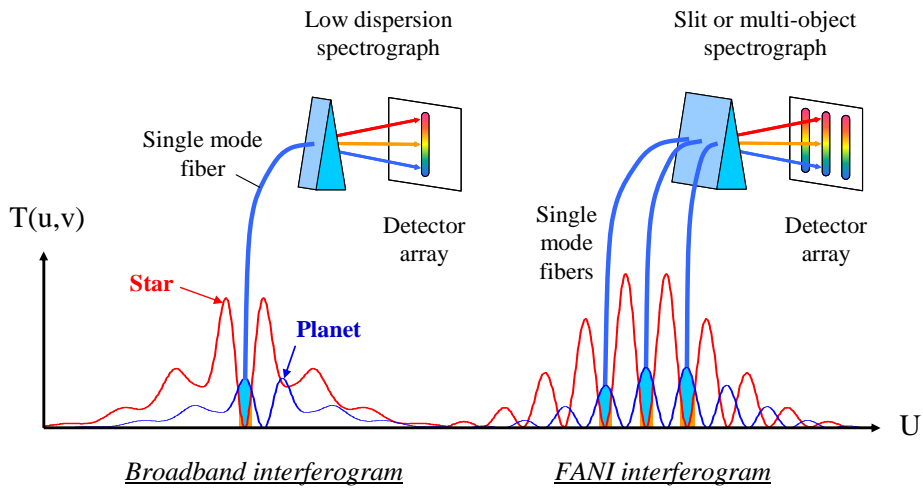

Figure 9: Illustrating the potential SNR gain inherent to an achromatized fringe pattern.

## 5  CONCLUSION

We described the concept of FANI, the Fully achromatic nulling interferometer enabling the creation of wavelength-independent fringe patterns – an idea that can also be envisioned as an achromatic variant of Young's double slit experiment. Practically, this is achieved with the help of dispersive optics located at an intermediate image plane of the interferometer system. After explaining the general principle and illustrating it with numerically simulated fringe patterns, we have described a preliminary optical design based on a grism mirror, and demonstrated that its alignment and manufacturing tolerances of the dispersive element are affordable. As usual in nulling interferometry the main technical difficulties should be concentrated on the APS specifications. It must be highlighted, however, that FANI's concept is not limited to the field of extra-solar planet detection and could probably benefit advantageously to classical stellar interferometers for broadband imaging of extended celestial sources. In both cases the method should provide a significant SNR gain to scientific observations.

# APPENDIX. COORDINATE SYSTEMS AND MATHEMATICAL RELATIONS

Let us consider a multi-aperture optical system composed of N collecting telescopes and N recombining sub-apertures. The three main employed coordinate systems are depicted in Figure 10:

- An on-sky angular coordinates system (U,V) where celestial objects are defined by angular coordinates u and v,
- An entrance pupil reference frame (O,X,Y,Z), where OZ is the main optical axis and OXY the input pupil plane,
- An exit pupil reference frame (O',X',Y',Z) where O'X'Y' is the output pupil plane,

Let us further assume that:

1) For all indices n comprised between 1 and N, the centre $P_n$ of the $n^{th}$ collecting aperture is optically conjugated with the centre $P'_n$ of its associated output aperture without pupil aberration.

2) All collecting apertures have identical diameters D and consequently all recombining apertures share the same diameter D'. Practically, it means that all collecting telescopes and optical trains conveying the beams from the entrance to the exit apertures are identical, which is very often the case in contemporary interferometer facilities.

Let us finally define the following parameters (bold characters denoting vectors):

| | |
|---|---|
| **s** | A unit vector of direction cosine ≈ (u,v,1) directed at any point in the sky (corresponding to any point M" in the image plane), where angular coordinates u and v are considered as first-order quantities |
| **s_O** | A unit vector of direction cosine ≈ ($u_O$,$v_O$,1) pointed at a given sky object (or an elementary angular area of it) |
| O(**s_O**) | The angular brightness distribution of an extended sky object |
| $\Omega_O$, $d\Omega_O$ | The total observed FoV in terms of solid angle, and its differentiating element |
| $PSF_T$(**s**) | The PSF of an individual collecting telescope, being projected back onto the sky. For an unobstructed pupil of diameter D, this would be the classical Airy distribution equal to $|2J_1(\rho)/\rho|^2$, where $\rho = k\,D\,\|\mathbf{s}\|/2$ and $J_1$ is the type-J Bessel function at the first order |
| k | The wavenumber $2\pi/\lambda$ of the electro-magnetic field assumed to be monochromatic, and where $\lambda$ is its wavelength |
| $a_n$ | The amplitude transmission factor of the $n^{th}$ interferometer arm ($1 \leq n \leq N$) |
| $\varphi_n$ | A phase-shift introduced along the $n^{th}$ interferometer arm for Optical Path Differences (OPD) compensation or nulling purposes ($1 \leq n \leq N$) |
| **OP_n** | A vector of coordinates ($x_n$,$y_n$,0) defining the center $P_n$ of the $n^{th}$ sub-pupil in the entrance pupil plane P ($1 \leq n \leq N$) |
| B | The maximal baseline between any couple (n,n') of telescopes ($1 \leq n$ and $n' \leq N$) |
| **O'P'_n** | A vector of coordinates ($x'_n$,$y'_n$,0) defining the center $P'_n$ of the $n^{th}$ sub-pupil in the exit pupil plane P' ($1 \leq n \leq N$) |
| B' | The maximal baseline between any couple (n,n') of exit sub-apertures ($1 \leq n$ and $n' \leq N$) |
| m | The optical compression factor of the system, equal to $m = D'/D = F_C/F$ where F and $F_C$ respectively are the focal length of the collecting telescopes and of the relay optics (see Figure 2). |

Hence the image I(**s**) formed by the multi-aperture optical system and projected back onto the sky writes in a first-order approximation:

$$I(\mathbf{s}) = \iint_{\mathbf{s_O} \in \Omega_O} O(\mathbf{s_O})\,PSF_T(\mathbf{s} - \mathbf{s_O}) \left| \sum_{n=1}^{N} a_n \exp[i\varphi_n] \exp[i k \xi(\mathbf{s_O},\mathbf{s})] \right|^2 d\Omega_O, \tag{A1}$$

with function ξ(s$_O$,s) being the OPD term:
$$\xi(\mathbf{s_O},\mathbf{s}) = \mathbf{s_O}\,\mathbf{OP_n} - \mathbf{s}\,\mathbf{O'P'_n}/m. \tag{A2}$$

The "generalized PSF" of this optical system is obtained by replacing the object brightness function O(s$_O$) in the integral with the impulse Dirac distribution δ(s-s$_O$) in Eq. A1:

$$\mathrm{PSF}_G(\mathbf{s},\mathbf{s_O}) = \mathrm{PSF}_T(\mathbf{s\text{-}s_O}) \left| \sum_{n=1}^{N} a_n \exp[i\varphi_n] \exp[ik\,\xi(\mathbf{s_O},\mathbf{s})] \right|^2. \tag{A3}$$

PSF$_G$(s,s$_O$) presents the particularity of constantly varying with s$_O$, the angular location of the sky object in the instrument Field of view (FoV). Assuming that all sub-apertures are circular and diffraction-limited and using the here above notations, Eq. A3 becomes:

$$\mathrm{PSF}_G(u,v,u_O,v_O) = \left| \frac{2J_1(\pi D r/\lambda)}{\pi D r/\lambda} \right|^2 \left| \sum_{n=1}^{N} a_n \exp[i\varphi_n] \exp\left[i\frac{2\pi}{\lambda}\left(u_0 x_n + v_0 y_n - \frac{u x'_n + v y'_n}{m}\right)\right] \right|^2, \tag{A4}$$

$$\text{where:} \quad r = \sqrt{(u-u_O)^2 + (v-v_O)^2}.$$

**Fizeau interferometers**

This study mainly deals with the special case of Fizeau interferometers, that is the unique configuration where a multi-aperture optical system forms "direct" images. In that case the output pupil of the interferometer is a scaled replica of its entrance pupil, i.e. all the entrance and exit sub-apertures as well as their relative arrangement are perfectly homothetic. Mathematically this condition implies that:

$$\mathbf{O'P'_n} = m\,\mathbf{OP_n}, \qquad (1 \le n \le N). \tag{A5}$$

also being known as the "golden rule of interferometry" [17]. Then Eq. A3 reduces to:

$$\mathrm{PSF}(\mathbf{s}) = \mathrm{PSF}_T(\mathbf{s}) \left| \sum_{n=1}^{N} a_n \exp[i\varphi_n] \exp[i\,k\,\mathbf{s}\,\mathbf{OP_n}] \right|^2, \tag{A6}$$

and the PSF is invariant in the whole FoV as in conventional Fourier optics.

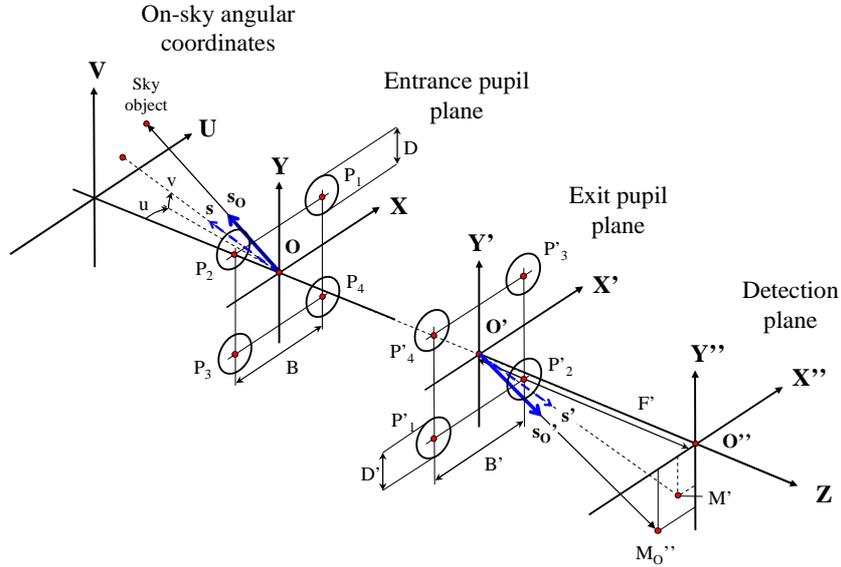

Figure 10: Sketch of the employed coordinate systems.